\documentclass[a4paper,11pt]{article}
\usepackage{jheppub} 
\usepackage{lineno}
\usepackage{slashed}

\title{\boldmath Radiative Corrections to the Muon Polarization in the Semileptonic Decay of a Neutral Kaon}

\author{J. Vieyra,  A. Martínez, M. Neri and A. Hernández-Galeana}
\affiliation{Departamento de Física, Escuela Superior de Física y Matemáticas del Instituto Politécnico Nacional, Apartado Postal 75-702,\\
Ciudad de México 07738, México }

\emailAdd{jmartinezvi@ipn.mx (corresponding author)}

\abstract{A model-independent expression for the Dalitz plot of the semileptonic decays of a neutral kaon $K_{\mu 3}^0$, including radiative corrections to order $\mathcal{O}[(\alpha / \pi )(q/M_1)]$, where $q$ is the four-momentum transfer and $M_1$ is the mass of the decaying kaon, is presented. In this paper the emitted muon is considered to be polarized so the analysis is centered on numerically evaluating the radiative corrections to the longitudinal, transverse, and normal polarization muon components. The model dependence of radiative corrections is kept in general form within this approximation, which is useful for model-independent experimental analyses. The final expressions, with the triple integration of the bremsstrahlung photon variables are ready to be performed numerically. The radiative corrections to the components of the muon polarization are found to be very small compared to their respective uncorrected values.}

\begin{document}
\maketitle
\flushbottom

\section{Introduction}
\label{sec:intro}

Kaon physics has played an crucial role in tests of $CP$ violation. The symmetry transformation $CPT$ is the combination of the three symmetry transformations $C$, $P$ and $T$, These transformations interchange particles and antiparticles, reverse spatial coordinates ($\mathbf{x} \rightarrow -\mathbf{x}$)  and  time ($t \rightarrow -t$), respectively.. According to the $CPT$ theorem, a Lorentz-invariant field theory remains unchanged under this transformation. Whereas $C$ and $P$ are maximally violated separately in weak interactions, $CP$ (and $T$) are nearly conserved. However, $CP$ violation was observed in the $K^0$ \cite{1964} and $B^0$ \cite{2001} systems in 1964 and 2001 respectively. This implies the existence of $T$ violation. Direct $CP$ violation can be detected through a difference in the normalized decay amplitudes for $K_L \rightarrow \pi^+ \pi^-$ with respect to $K_L \rightarrow \pi^0 \pi^0$. The search for $T$ violation, initially proposed by Sakurai, focuses on the transverse muon polarization $P_T$ in the decay $K^+ \rightarrow \pi^0 \mu^+ \nu$ (denoted as $K^+_{\mu 3}$) \cite{sakurai}. $P_T$ is a T-odd observable representing the polarization component perpendicular to the decay plane. It is defined as the correlation between the momentum vectors $\pi^0$ and $\mu^+$ and the spin vector $\mu^+$ .  Detecting a non-zero value of $P_T$ between $10^{-5}$ and $10^{-3}$, would strongly indicate a violation of time reversal invariance, while the Standard Model (SM) predicts $P_T$ values around the order of $10^{-7}$. On the experimental front, the KEK-E246 collaboration has placed an upper limit of $|P_{T}|<0.0050$ at a $90\%$ confidence level for charged kaons \cite{kek}. However, for neutral kaons, no experimental data have been reported since 1980 \cite{morse}, the value reported here is $P_T=0.0017 \pm 0.063$.

Furthermore, on the other hand, current precise measurements of some elements of the Cabibbo-Kobayashi-Maskawa matrix (CKM) such as $|V_{us}|$, are derived primarily from the analysis of kaon semileptonic $(K_{\ell3})$ decays and leptonic $(K_{\ell2})$ decays. Additionally, the most precise measure of $|V_{ud}|$ arises from the analysis of the superallowed $0^+ \rightarrow 0^+$ Fermi transitions and from baryon semileptonic decays.  The precision of this measurements, from a theoretical point of view, resides in the assumptions about the form factors, momentum transfer dependence of the form factors and radiative corrections (RC) to the integrated observables.

There are various works dealing with the radiative corrections to $K_{\ell3}$ decays, each one from a different approach. Some of the earliest attempts can be found in the works by Ginsberg \cite{gin1,gin2,gin3,gin4}, Becherrawy \cite{beche}, García and Maya \cite{maya}, and more recently by Cirigliano \textit{et al}. \cite{ciri1,ciri2,ciri3} and Seng \textit{et al}. \cite{seng1,seng2,seng3}. Ginsberg studied th RC to the lepton spectrum, DP, and decay rates of $K_{\ell3}$ decays, by assuming a phenomenological weak $K-\pi$ vertex. Becherrawy employs a model of strong interactions, García and Maya extended the Sirlin methodology \cite{sirlin} to $M_{\ell3}$, which was introduced to calculate radiative corrections to the charged lepton spectrum in neutron beta decay. On the other hand, Cirigliano \textit{et al}. implemented chiral perturbation theory taking into account virtual photons and leptons. Senge \textit{et al}. use a hybrid analysis based on current algebra, chiral perturbation theory and lattice QCD. 

In this paper, we reexamine the calculation of radiative corrections to $K_{\ell3}^0$ decays up to the order $\mathcal{O}[(\alpha / \pi )(q/M_1)]$, where $q$ is the four-momentum transfer and $M_1$ is the mass of the decaying kaon. Our analysis builds on earlier works, for charged kaons the so-called three-body and four-body region were analysed in Refs. \cite{2011} and \cite{2012}, whereas for neutral kaons both regions have been treated in Refs. \cite{2015} and \cite{2016} . Particularly in Ref \cite{2020}, a more complete analysis of the charged kaon has been performed, considering the muon polarization, but unlike it, here we will focus on giving theoretical expressions to evaluate the RC to the three components of the muon polarization vector for $K_{\ell3}^0$ decays.

This paper is organized in the following way. In Sect. \ref{section:over} we give an overview of kaon semileptonic decays. In Sect. \ref{section:virtual} we discuss the analysis of virtual RC, dealing with the spin-dependent part. In Sect. \ref{section:brems} the analysis of bremsstrahlung RC is discussed in the three body region (TBR), and the integrals are numerically evaluated. Later in Sect. \ref{section:dp} the complete DP is presented. Finally in Sect. \ref{section:pola} we present the muon polarization vector including RC and its components are numerically evaluated at several points of the allowed kinematical region as well. The findings are presented and discussed in Sect. \ref{section:con}. 

\section{An Overview of Kaon Semileptonic Decays}
\label{section:over}

\noindent
For definiteness, the semileptonic decay of a neutral kaon, hereafter referred to as $K_{\ell 3}^0$, is represented by

\begin{equation}
    K^0(p_1)\rightarrow \pi^-(p_2)+\ell^+(l)+\nu_{\mu}(p_{\nu}),
    \label{proceso}
\end{equation}

\noindent
where the four-momenta of $K^0$, $\pi^-$, $\ell^+$ and $\nu_\mu$ are $p_1=(E_1,\mathbf{p}_1)$, $p_2=(E_2,\mathbf{p}_2)$, $l=(E,\mathbf{l})$ y $p_\nu=(E_\nu^0,\mathbf{p}_\nu)$, and their masses are $M_1$, $M_2$, $m$ y $m_\nu$, respectively. The case studied in this work is $\ell=\mu$. The chosen reference system used is the rest frame of the $K^0$, which serves as the point of reference for all noncovariant expressions. Within this framework, terms such as $p_2$, $l$, or $p_\nu$ will also denote the magnitudes of their respective three-momenta, unless explicitly stated otherwise.  Furthermore, we indicate the direction of any generic three-vector $\mathbf{p}$ using the unit vector denoted as $\mathbf{\hat{p}}$.

The uncorrected transition amplitude (i.e. the amplitude without RC) for our decay (\ref{proceso}), using $V-A$ theory is given by 

\begin{equation}
    M_0=C_K\frac{G_F}{\sqrt{2}}V_{us}^*W_{\alpha}(p_1,p_2) \left[\bar{u}_\nu(p_\nu)O_\alpha v_\mu(l)\right],
\end{equation}

\noindent
where $C_K$ is the Clebsch-Gordan coefficient that is equal to 1, in contrast with the  charged counterpart $K^\pm$, where is equal to $1/\sqrt{2}$, $G_F$ is the Fermi constant, $V_{us}$ is the element of the Cabibbo-Kobayashi-Masakawa (CKM) matrix and $W_\alpha(p_1,p_2)=f_+(q^{2})(p_1+p_2)_\alpha+f_-(q^{2})(p_1-p_2)_\alpha$, where $f_\pm(q^2)$ are the form factors that depend on the four-momentum transfer $q\equiv p_1-p_2$. Here $v_\mu$ and $u_\nu$ denote the Dirac spinors for the muon and neutrino respectively and $O_\alpha\equiv \gamma_\alpha(1+\gamma_5)$. The metric and the conventions on the algebra of $\gamma$ matrices are specified in \cite{2011}.

\noindent
In the $K_{\mu 3}$ decay analyses it is a standard procedure to assume a linear dependence of $f_\pm$ on $q^2$, i.e.,

\begin{equation}
    f_\pm(q^2)=f_\pm(0)\left[1+\lambda_\pm \frac{q^2}{M_2^2} \right].
\end{equation}

\noindent
and most data are characterized by a constant $f_-$ \cite{pdg}.

\noindent
An alternative way to parameterize the form factors involves introducing the ratio

\begin{equation}
    \xi(q^2)\equiv \frac{f_-(q^2)}{f_+(q^2)}.
\end{equation}

\noindent
Hence, the pertinent parameters are $\lambda_+$ and $\xi(0)$. 

In recent analyses, the form factors $f_+$ and $f_0$ are used instead, these are related by 
\cite{pdg}

\begin{equation}
    f_0(q^2)=f_+(q^2)+\frac{q^2}{M_1^2-M_2^2}f_-(q^2).
\end{equation}

\noindent
For a linear dependence of $f_+$ in $q^2$ and $f_-$ to be constant, $f_0(q^2)$ can be written as

\begin{equation}
    f_0(q^2)=f_0(0)\left[1+\lambda_0\frac{q^2}{M_2^2} \right],
\end{equation}

\noindent
under the same assumptions, $\xi(q^2)$ can be expressed as

\begin{equation}
    \xi(q^2)=\frac{M_1^2-M_2^2}{M_2^2}\frac{\lambda_0-\lambda_+}{1+\lambda_+\frac{q^2}{M_2^2}}=\xi(0)\left[1+\lambda_+\frac{q^2}{M_2^2} \right]^{-1}.
    \label{xi}
\end{equation}

\noindent
To analyze the muon polarization, the transition amplitude can be written as

\begin{equation}
    M_0 =C_K\frac{G_F}{\sqrt{2}}V_{us}^{*}f_+(q^2)\left[ 2{p_1}_\alpha -\left[1-\xi(q^2) \right]q_\alpha\right]\left[\bar{u}_\nu(p_\nu)O_\alpha v_\mu(l)\right].
\end{equation}

\noindent
The polarization of the emitted muon can be considered using the spin projection operator

\begin{equation}
    \Sigma (s) =\frac{1-\gamma_5 \slashed{s}}{2},
    \label{proyector}
\end{equation}

\noindent
where $s\cdot s=s_0^2-\mathbf{s}\cdot \mathbf{s}=-1$ and $s\cdot l=0$. In the muon's rest frame, $s$ reduces to the unit vector $\hat{\mathbf{s}}_R$, which indicates the spin direction. The observable effects of spin polarization can be analyzed replacing

\begin{equation}
    v_\mu(l)\rightarrow \Sigma(s)v_\mu(l),
    \label{sust}
\end{equation}

\noindent
in the corresponding muon's spinor in the transition amplitude $M_0$.

\noindent
Squaring the decay amplitude and performing a summation over the spins in the final state, yields to

\begin{equation}
    \sum_{\text{spins}} |M_0|^2=\frac{1}{2}  \sum_{\text{spins}}  |M_0'|^2 -\frac{1}{2}  \sum_{\text{spins}}  |M_0^{(s)}|^2,
\end{equation}

\noindent
where the first and second parts contain the spin-independent and spin-dependent contributions to $\sum|M_0|^2$.

\noindent
The uncorrected differential decay rate for (\ref{proceso}) is given by

\begin{equation}
 d\Gamma_0 =\frac{1}{2M_1}\frac{d^3p_2}{2E_2(2\pi)^3}\frac{m}{E}\frac{d^3l}
 {(2\pi)^3}\frac{m_\nu}{E_\nu^0}\frac{d^3p_\nu}{(2\pi)^3} (2\pi)^4 \delta^4(p_1-p_2-l-p_\nu)\sum_{\text{spins}} |M_0|^2.
\end{equation}

\noindent
 $d\Gamma_0$ in the kaon rest system, leaving  $E$ and $E_2$ as independent variables, yields the Dalitz plot. For the integrals over the angular variables of the muon and pion, the coordinate axes are oriented in such a way that $\mu^+$ is emitted along the $z^+$ axis and $\pi^-$ be emitted in the first quadrant of the $(x,z)$ plane. The last nontrivial integral is over the polar angle of $\pi^-$, $\theta_2$, namely

\begin{equation}
    d\Gamma_0=\frac{1}{(2\pi)^3}\frac{m m_\nu}{2M_1}dEdE_2\int_{-1}^{1}dy\delta(y-y_0)\sum_{\text{spins}} |M_0|^2,
\end{equation}

\noindent
where 

\begin{equation}
    y_0=\frac{(E_\nu^0)^2-p_2^2-l^2}{2p_2l},
\end{equation}

\noindent 
and $y=\cos \theta_2$ is the cosine of the angle between $\mathbf{p}_2$ and $\mathbf{l}$. Additionally, energy conservation yields $E_\nu^0=M_1-E_2-E$.

\subsection{Muon Polarization}

\noindent
At this step, the total decay rate can be obtained, which can be separated into the corresponding spin-independent and spin-dependent parts. The total decay rate is described by

\begin{equation}
    d\Gamma_0(K^0_{\mu 3})=\frac{1}{2}d\Gamma_0'+\frac{1}{2}d\Gamma_0^{(s)}.
\end{equation}

The muon spin $s$ in the kaon rest frame is related to its in the muon rest frame $\hat{\mathbf{s}}_R$ by 

\begin{equation}
s_0=\frac{1}{m}\hat{\mathbf{s}}_R\cdot \mathbf{l}, \quad s_\parallel=\frac{E}{m}(\hat{\mathbf{s}}_R\cdot \hat{\mathbf{l}})\hat{\mathbf{l}}, \quad s_\perp=\hat{\mathbf{s}}_R-(\hat{\mathbf{s}}_R\cdot \hat{\mathbf{l}})\hat{\mathbf{l}},
\end{equation}

\noindent
this way, if $a$ is an arbitrary four-vector, then

\begin{equation}
s\cdot a= \hat{\mathbf{s}}_R\cdot \biggl[\frac{\mathbf{l}}{m}\biggl( a_0-\frac{\mathbf{a}\cdot \mathbf{l}}{E+m}\biggr)-\mathbf{a} \biggr].
\label{trans}
\end{equation}

\noindent

The decay plane is generated by the vectors $\mathbf{l}$ and $\mathbf{p}_2$, so that, the three components of the uncorrected muon polarization $\mathbf{P}_0$ are given by

\begin{equation}
P_{L0}=\mathbf{P}_0 \cdot \hat{\mathbf{\epsilon}}_L, \quad P_{T0}=\mathbf{P}_0 \cdot \hat{\mathbf{\epsilon}}_T, \quad P_{N0}=\mathbf{P}_0 \cdot \hat{\mathbf{\epsilon}}_N,
\label{uncpol}
\end{equation}

\noindent
namely, the longitudinal $P_{L0}$, transverse $P_{T0}$ and normal $P_{N0}$, here the subscript $0$ denotes an uncorrected quantity. Additionally the vectors  $(\hat{\mathbf{\epsilon}}_L,\hat{\mathbf{\epsilon}}_T,\hat{\mathbf{\epsilon}}_N)$ form an orthonormal basis and are defined as

\begin{equation}
\hat{\mathbf{\epsilon}}_L=\frac{\mathbf{l}}{|\mathbf{l}|}, \quad \hat{\mathbf{\epsilon}}_T=\frac{\mathbf{p}_2 \times \mathbf{l}}{|\mathbf{p}_2 \times \mathbf{l}|}, \quad \hat{\mathbf{\epsilon}}_N=\hat{\mathbf{\epsilon}}_L \times \hat{\mathbf{\epsilon}}_T.
\end{equation}


\noindent
Note that $\hat{\epsilon}_L$ and $\mathbf{l}$ are parallel, $\hat{\epsilon}_T$ is perpendicular to the decay plane and $\hat{\epsilon}_N$ is normal to both $\hat{\epsilon}_L$ and $\hat{\epsilon}_T$ in the decay plane.

\noindent
The spin-independent contribution to the uncorrected differential decay rate is

\begin{equation}
d\Gamma_0'(E,E_2)=a_0'd\Omega',
\end{equation}


\noindent
where $a_0'$ reads

\begin{equation}
a_0'=2M_1EE_\nu^0-M_1^2(E_{2m}-E_2)-m^2E_\nu^0\text{Re}[1-\xi(q^2)]
+\frac{1}{4}m^2(E_{2m}-E_2)|1-\xi(q^2)|^2,
\label{a0p}
\end{equation}

\noindent
and

\begin{equation}
d\Omega'=\frac{C_k^2G_F^2|V_{us}|^2|f_+|^2}{4\pi^3} dEdE_2.
\end{equation}

\noindent
In the same way, the spin-dependent contribution can be written as

\begin{equation}
d\Gamma^{(s)}_0=\hat{\mathbf{s}}_R\cdot \mathbf	{a}_0^{(s)}d\Omega',
\end{equation}

\noindent
where $\mathbf{a}_0^{(s)}$ in terms of the orthonormal basis reads

\begin{equation}
\mathbf{a}_0^{(s)}=\Lambda_{L0}\hat{\epsilon}_L+\Lambda_{N0}\hat{\epsilon}_N+\Lambda_{T0}\hat{\epsilon}_T,
\label{coef}
\end{equation}

\noindent
where, the different $\Lambda_{X0}$ functions $(X=L,T,N)$ are defined as

\begin{equation}
\Lambda_{L0}=M_1\left[lE_\nu^0-E(l+p_2y_0)\right]+m^2(l+p_2y_0)\text{Re}[1-\xi(q^2)]-\frac{m^2}{4M_1}|1-\xi(q^2)|^2\left[lE_\nu^0+E(l+p_2y_0) \right],
\end{equation}

\begin{equation}
\Lambda_{T0}=mp_2l\sqrt{1-y_0}\text{Im}\xi(q^2),
\end{equation}

\noindent
and

\begin{equation}
\Lambda_{N0}=mp_2\sqrt{1-y_0}\left[M_1-E\text{Re}[1-\xi(q^2)]+\frac{m^2}{4M_1}|1-\xi(q^2)|^2 \right].
\end{equation}

\noindent
The differential decay rate can also be expressed as follows

\begin{equation}
d\Gamma_0=\frac{1}{2}d\Gamma_0'(1+\hat{\mathbf{s}}_R\cdot \mathbf{P}_0),
\end{equation}

\noindent
where $\mathbf{P}_0$ is the muon polarization vector defined in Eq. (\ref{uncpol}), which explicitly reads

\begin{equation}
\mathbf{P}_0=P_{L0}\hat{\epsilon}_L+P_{T0}\hat{\epsilon}_T+P_{N0}\hat{\epsilon}_N.
\label{polarizacion0}
\end{equation}

\noindent
The uncorrected components of the muon polarization can be written as

\begin{equation}
P_{X0}=\frac{\Lambda_{K0}}{a_0'}.
\end{equation}

\noindent
The components of $\mathbf{P}_0$ are listed in Table \ref{tabla0}, where the parameter $\xi(q^2)$ was used as described in Eq. (\ref{xi}).

\noindent
To close this section, it should be noted that $|\mathbf{P}_0|=1$ at each point in the kinematical region of the Dalitz plot.

\begin{table*}[ht]
\centering
\resizebox{15cm}{!} {
\begin{tabular}{c|rrrrrrrrrr}
\hline
 $E_2$\textbackslash$E$ &$0.1123$&$0.1258$&$0.1393$&$0.1528$&$0.1663$&$0.1797$&$0.1932$&$0.2067$&$0.2202$&$0.2337$\\ \hline
 (a)\\
        $0.2512$ & ~ & ~ & ~ & ~ & ~ & $0.9969$ & $0.9948$ & $0.9928$ & $0.9893$ & $0.9718$   \\ 
        
        $0.2395$ & ~ & $0.9719$ & $0.9593$ & $0.9613$ & $0.9650$ & $0.9678$ & $0.9689$ & $0.9671$ & $0.9578$ & $0.8876$\\ 
        
        $0.2277$ &  $0.7686$ & $0.8224$ & $0.8712$ & $0.9016$ & $0.9205$ & $0.9318$ & $0.9370$ & $0.9352$ & $0.9170$ & $0.7406$  \\ 
        
        $0.2160$ & $0.2232$ & $0.6110$ & $0.7518$ & $0.8230$ & $0.8631$ & $0.8860$ & $0.8967$ & $0.8944$ & $0.8622$ & $0.4221$  \\ 
        
        $0.2042$ & $-0.6918$ & $0.2893$ & $0.5813$ & $0.7153$ & $0.7867$ & $0.8260$ & $0.8442$ & $0.8405$ & $0.7851$ & $-0.7753$ \\ 
        
        $0.1924$ & ~ & $-0.2585$ & $0.3178$ & $0.5589$ & $0.6800$ & $0.7441$ & $0.7730$ & $0.7665$ & $0.6689$ &   \\ 
        
        $0.1807$ & ~ & ~ & $-0.1423$ & $0.3112$ & $0.5207$ & $0.6260$ & $0.6715$ & $0.6582$ & $0.4739$ &   \\ 
        
        $0.1689$ & ~ & ~ & ~ & $-0.1398$ & $0.2576$ & $0.4409$ & $0.5150$ & $0.4856$ & $0.0808$ &   \\ 
        
        $0.1572$ & ~ & ~ & ~ & ~ & $-0.2591$ & $0.1097$ & $0.2430$ & $0.1674$ & ~ &   \\ 
        
        $0.1454$ & ~ & ~ & ~ & ~ & ~ & $-0.6518$ & $-0.3465$ & $-0.6133$ & ~ &   \\
 
 (b)\\
     $0.2512$ & ~ & ~ & ~ & ~ & ~ & $-0.0267$ & $-0.0409$ & $-0.0557$ & $-0.0782$ & $-0.1453$\\ 
        
        $0.2395$ & ~ & $-0.0313$ & $-0.0520$ & $-0.0645$ & $-0.0749$ & $-0.0857 $& $-0.0988$ & $-0.1179$ & $-0.1538$ & $-0.2833$ \\ 
        
        $0.2277$ & $-0.0456$ & $-0.0755$ & $-0.0903$ & $-0.1013$ & $-0.1116$ & $-0.1235$ & $-0.1393$ & $-0.1640$ & $-0.2131$ & $-0.4123$  \\ 
        
        $0.2160$ &  $-0.0694$ & $-0.1049$ & $-0.1212$ & $-0.1328$ & $-0.1441$ & $-0.1575$ & $-0.1763$ & $-0.2068$ & $-0.2701$ & $-0.5552$  \\ 
        
        $0.2042$ & $-0.0514$ & $-0.1268$ & $-0.1495$ & $-0.1633$ & $-0.1759$ & $-0.1912$ & $-0.2132$ & $-0.2501$ & $-0.3297$ & $-0.3860$  \\ 
        
        $0.1924$ & ~ & $-0.1279$ & $-0.1741$ & $-0.1936$ & $-0.2087$ & $-0.2264$ & $-0.2520$ & $-0.2961$ & $-0.3951$  &   \\ 
        
        $0.1807$ & ~ & ~ & $-0.1816$ & $-0.2217$ & $-0.2428$ & $-0.2640$ & $-0.2940$ & $-0.3466$ & $-0.4672$ &   \\ 
        
        $0.1689$ & ~ & ~ & ~ & $-0.2308$ & $-0.2746$ & $-0.3035$ & $-0.3397$ & $-0.4019$ & $-0.5280$ &   \\ 
        
        $0.1572$ & ~ & ~ & ~ & ~ & $-0.2742$ & $-0.3358$ & $-0.3839$ & $-0.4526$  & ~ &   \\ 
        
        $0.1454$ & ~ & ~ & ~ & ~ & ~ & $-0.2559$ & $-0.3709$ & $-0.3621$ & ~ &   \\
 
 (c)\\
     $0.2512$ & ~ & ~ & ~ & ~ & ~ & $-0.0782$ & $-0.1022$ & $-0.1199$ & $-0.1459$ & $-0.2357$ \\ 
        
        $0.2395$ & ~ & $-0.2356$ & $-0.2824$ & $-0.2754$ & $-0.2621$ & $-0.2515$ & $-0.2473$ & $-0.2543$ & $-0.2875$ & $-0.4607$ \\ 
        
        $0.2277$ &  $-0.6397$ & $-0.5689$ & $-0.4909$ & $-0.4326$ & $-0.3908$ & $-0.3630$ & $-0.3492$ & $-0.3542$ & $-0.3989$ & $-0.6719$\\ 
        
        $0.2160$ &  $-0.9748$ & $-0.7916$ & $-0.6593$ & $-0.5680$ & $-0.5050$ & $-0.4637$ & $-0.4426$ & $-0.4473$ & $-0.5065$ & $-0.9065$ \\ 
        
        $0.2042$ & $-0.7221$ & $-0.9572$ & $-0.8137$ & $-0.6988$ & $-0.6173$ & $-0.5637$ & $-0.5361$ & $-0.5417$ & $-0.6193$ & $-0.6316$ \\ 
        
        $0.1924$ & ~ & $-0.9660$ & $-0.9482$ & $-0.8292$ & $-0.7332$ & $-0.6680$ & $-0.6344$ & $-0.6423$ & $-0.7434$ &   \\ 
        
        $0.1807$ & ~ & ~ & $-0.9898$ & $-0.9503$ & $-0.8537$ & $-0.7798$ & $-0.7410$ & $-0.7528$ & $-0.8806$ &   \\ 
        
        $0.1689$ & ~ & ~ & ~ & $-0.9902$ & $-0.9660$ & $-0.8975$ & $-0.8572$ & $-0.8742$ & $-0.9967$ &   \\ 
        
        $0.1572$ & ~ & ~ & ~ & ~ & $-0.9658$ & $-0.9940$ & $-0.9700$ & $-0.9859$ & ~ &   \\ 
        
        $0.1454$ & ~ & ~ & ~ & ~ & ~ & $-0.7583$ & $-0.9381$ & $-0.7898$ & ~ &   \\
    \hline    
        
\end{tabular}
}

\caption{\label{tabla0}Values of the components of the uncorrected muon polarization $\mathbf{P}_0$. Eq. (\ref{xi}), in the TBR of the process $K_{\mu 3}^0$. The entries correspond to (a)$P_{L0}$, 
 (b)$P_{T0}\times 10^2$, and (c)$P_{N0}$. The energies $E$ and $E_2$ are given in GeV. For definiteness, $\text{Re}\xi(0)=-0.151$ and $\text{Im}\xi(0)=-0.007$ are used.\label{tab:0}}
\end{table*}

\section{Virtual radiative corrections}
\label{section:virtual}

There are two types of RC need consideration: virtual RC and bremsstrahlung RC. Virtual RC can be split into two components: a model-independent that is finite, and a model-dependent that encloses the effects of strong interactions and the intermediate vector boson. Bremsstrahlung RC can be computed to the same order of approximation using the Low theorem \cite{low}.

\noindent
The model-independent transition amplitude with virtual RC to order $\mathcal{O}[(\alpha / \pi)(q/ M_1)]$ is given in Eq. (12) of Ref. \cite{2015}. It's given by

\begin{equation}
\text{M}_V=\text{M}_0'\left[1+\frac{\alpha}{2\pi}\Phi_n(E,E_2)\right]-\frac{\alpha}{2\pi}\text{M}_{p_2}\Phi_n'(E,E_2),
\label{avirt}
\end{equation}

\noindent
where the amplitude $\text{M}_{p_2}$ and the functions $\Phi_n(E,E_2)$ and $\Phi_n'(E,E_2)$, can be found in Eqs. (14), (15) and (16) of \cite{2015}. This terms make the difference between the neutral and charged counterpart, Eq. (36) of \cite{2020}.

\noindent
The spin-independent part of the quantity $\sum_s |\text{M}_V|^2$ has been evaluated in Ref. \cite{2015}, whereas the spin-dependent part is evaluated here. The analysis of the case of an emitted polarized muon can be worked out as in the previous section by introducing again the spin projection operator (\ref{proyector}) in the corresponding spinor of the muon in Eq. (\ref{avirt}). In this way, the differential decay rate of $K_{\ell3}^0$ decays for polarized emitted muons, with virtual RC, can be written as

\begin{equation}
    d\Gamma_V^{(s)}=d\Omega' \hat{\mathbf{s}}_R \cdot \left[\left(1+\frac{\alpha}{\pi} Re(\Phi_n) \right)\mathbf{a}_0^{(s)}+\frac{\alpha}{\pi}Re(\Phi'_n)\mathbf{a}_{V}^{(s)}  \right],
    \label{virtuals}
\end{equation}

\noindent
where $\mathbf{a}_0^{(s)}$ has been defined in Eq. (\ref{coef}) and

\begin{equation}
\mathbf{a}_{V}^{(s)}=\frac{EE_2-lp_2y_0}{m^2}\mathbf{a}_0^{(s)}-\left[\frac{lE_2-p_2Ey_0}{m^2} a_0'\right]\hat{\epsilon}_L+\left[\frac{p_2}{m}\sqrt{1-y_0^2}a_0' \right]\hat{\epsilon}_N ,
\end{equation}

\noindent
with $\beta=l/E$.

\section{Bremsstrahlung radiative corrections}
\label{section:brems}

A comprehensive analysis of RC to the Dalitz plot should consider the emission of an actual photon through the following process

\begin{equation}
K^0(p_1)\rightarrow \pi^-(p_2)+\mu^+(l)+\nu_{\mu}(p_{\nu})+\gamma(k),
\label{decay4}
\end{equation}

\noindent
where $\gamma$ denotes a photon with four-momentum $k=(\omega,\mathbf{k})$ and the neutrino four momentum is now $p_\nu=(E_\nu,\mathbf{p_\nu})$.

\noindent
According the Low theorem \cite{low}, the decay amplitude for process (\ref{decay4}) can be written as

\begin{equation}
M_B=\sum_{i=1}^4M_{B_i},
\end{equation}

\noindent
where the pieces of this equation read

\begin{equation}
\text{M}_{B_1}=-e\text{M}_0\left[\frac{l\cdot \varepsilon}{l\cdot k}-\frac{p_2\cdot \varepsilon}{p_2\cdot k} \right],
\label{b1}
\end{equation}

\begin{equation}
\text{M}_{B_2}=-C_K\frac{eG_F}{\sqrt{2}}V_{us}\bar{u}_\nu\mathcal{O}^\alpha\frac{\slashed{k}\slashed{\varepsilon}}{2l\cdot k} \nu_\mu,
\label{b2}
\end{equation}

\begin{equation}
\text{M}_{B_3}=-C_K\frac{eG_F}{\sqrt{2}}V_{us}(f_+-f_-)\left[-\frac{p_2\cdot \varepsilon}{p_2\cdot k}k_\alpha+\varepsilon_\alpha \right]\bar{u}_\nu\mathcal{O}^\alpha \nu_\mu,
\label{b3}
\end{equation}

\noindent
and

\begin{equation}
\text{M}_{B_4}=-C_K\frac{eG_F}{\sqrt{2}}V_{us}\left[ \frac{p_1\cdot \varepsilon}{p_1\cdot k}q\cdot k-q\cdot \varepsilon \right]\frac{\partial W_\lambda}{\partial q^2} \bar{u}_\nu \mathcal{O}_\lambda \nu_l.
\end{equation}

\noindent

\noindent
It is important to mention that the amplitude $\text{M}_{B_1}$ is of order $\mathcal{O}(1/k)$ and contains the infrared divergence, whereas $\text{M}_{B_2}$ and $\text{M}_{B_3}$ are of order $\mathcal{O}(k^0)$. Furthermore, the contribution of $\text{M}_{B_4}$ will be neglected because produces terms of order $q^2/M_1^2$ to the decay rate, which are not needed in the present analysis. In contrast with the charged counterpart, $\text{M}_{B_1}$ and $\text{M}_{B_3}$ carry terms with $p_2\cdot k$ which leads to longer expressions in the differential decay rate.

\noindent
The differential decay rate for bremsstrahlung radiation can be derived using conventional methods, which is

\begin{equation}
d \Gamma_B=\frac{1}{(2\pi)^8}\frac{1}{2M_1}\frac{mm_\nu}{4E_2EE_\nu \omega}d^3p_2d^3ld^3p_\nu d^3k\delta^4(p_1-p_2-l-p_\nu-k) \sum_{\text{spins,pol.}} |M_B|^2, 
\end{equation}

\noindent
where, the observable effects of spin polarization can be analyzed by making the replacement indicated in Eq. (\ref{sust}), this enables us to once more separate the spin-independent component from the spin-dependent component as follows

\begin{equation}
   \sum_{\text{spins,pol.}} |\text{M}_B|^2=\frac{1}{2}\sum_{\text{spins,pol.}} |\text{M}_B'|^2-\frac{1}{2}\sum_{\text{spins,pol.}} |\text{M}_B^{(s)}|^2.
   \label{mbrems}
\end{equation}

\noindent
The first has been treated in Ref. \cite{2015}, whereas as in the previous section, the spin-dependent part is evaluated here.

\noindent
The orientation of the coordinate axes is such that the emission of the muon it's aligned with the $+z$ axis and the pion is emitted in the first or fourth quadrant of the $(y,z)$ plane. For definiteness, let $\hat{\mathbf{p}}_2 \cdot \hat{\mathbf{l}}=\cos \theta_2\equiv y$, $\hat{\mathbf{l}} \cdot \hat{\mathbf{k}}=\cos \theta_k\equiv x$ and $\hat{\mathbf{p}}_2 \cdot \hat{\mathbf{k}}=\cos \theta_2 \cos \theta_k+\sin \theta_2 \sin \phi_k$, where $\theta_k$ y $\phi_k$ are the photon's polar and azimuthal angles, and its energy is given by

\begin{equation}
    \omega=\frac{F}{2D},
\end{equation}

\noindent
with

\begin{equation}
    F=2p_2 l(y_0-y),
\end{equation}

\noindent
and

\begin{equation}
    D=E_\nu^0+lx+\mathbf{p}_2 \cdot \hat{\mathbf{k}}.
\end{equation}

\noindent
The Three Body Region TBR of the Dalitz Plot can be seen as the region where the three-body decays (\ref{proceso}) and (\ref{decay4}) overlap entirely. In the TBR the energies $E$ and $E_2$ are restricted to

\begin{equation}
    m \leq E \leq E_m, \quad E_2^{min} \leq E_2 \leq E_2^{max},
\end{equation}

\noindent
where

\begin{equation}
    E_m=\frac{M_1^2-M_2^2+m^2}{2M_1},
\end{equation}

\noindent
and

\begin{equation}
    E_2^{max,min}=\frac{1}{2}(M_1-E\pm l)+\frac{M_2^2}{(M_1-E\pm l)},
\end{equation}

\noindent
The variable $y$ is restricted to $-1\leq y \leq y_0$.

\noindent
Following the analysis of the previous section, $d\Gamma_B$ can also be separated as

\begin{equation}
    d\Gamma_B=\frac{1}{2}d\Gamma_B'+\frac{1}{2}d\Gamma_B^{(s)},
    \label{bremss}
\end{equation}

\noindent
where $d\Gamma_B'$ is the spin-independent bremsstrahlung contribution to the DP, which has been obtained in Ref. \cite{2015}, the latter $d\Gamma_B^{(s)}$ is the spin-dependent bremsstrahlung contribution to the DP, which constitutes the primary objective of this section.

\subsection{Bremsstrahlung RC in the TBR}

Following the procedure of Ref. \cite{2015} and performing some algebraic manipulations, the second term in Eq. (\ref{bremss}) can be written as

\begin{equation}
    d\Gamma_B^{(s)}=\frac{\alpha}{\pi}d\Omega' \hat{\mathbf{s}}_R \cdot \left[\mathbf{a}_0^{(s)}I_0(E,E_2)+\mathbf{a}_B^{(s)}  \right].
    \label{brems}
\end{equation}

\noindent
In Eq. (\ref{brems}) the first term contains the infrared divergence inside the quantity $I_0(E,E_2)$ given in Eq. (52) of Ref. \cite{2015}. The term $\mathbf{a}_0^{(s)}$ is given in Eq. (\ref{coef}). The second summand, $\mathbf{a}_B^{(s)}$, is originated by the convergent pieces of $\sum_s|\text{M}_B^{(s)}|^2$, given in Eq. (\ref{mbrems}), and can be written as

\begin{equation}
\mathbf{a}_B^{(s)}=\Lambda_{LB}\hat{\epsilon}_L+\Lambda_{NB}\hat{\epsilon}_N+\Lambda_{TB}\hat{\epsilon}_T.
\end{equation}

\noindent
The explicit forms of the $\Lambda_{XB}$ are quite large so for the purposes of this paper, they will be omitted and evaluated numerically instead.

\section{Complete Dalitz plot}
\label{section:dp}

The differential decay rate of $K^0_{\mu 3}$ decays in terms of the energies of the muon and the pion, that is, the DP, with nonzero muon's polarization including RC to order $\mathcal{O}(\alpha/\pi)(q/M_1)$ is given by

\begin{equation}
d\Gamma(K_{\mu 3}^0)=d\Gamma_V+d\Gamma_B=\frac{1}{2}\left[d\Gamma'_V+d\Gamma'_B \right]+\frac{1}{2}\left[d\Gamma_V^{(s)}+d\Gamma_B^{(s)} \right],
\label{total}
\end{equation}

\noindent
where the primed quantities represent the unpolarized case  analyzed in Ref. \cite{2015}, the remaining quantities, with the superscript $(s)$, are consequence of muon's polarization itself. $d\Gamma_V^{(s)}$ is given by Eq. (\ref{virtuals}) and the bremsstrahlung counterpart is given by Eq. (\ref{brems}).

\noindent
Eq. (\ref{total}) can be written as

\begin{equation}
    d\Gamma(K_{\mu 3}^0)=\frac{1}{2}d\Omega' \left[\left(a_0' +\frac{\alpha}{\pi}a\right)+ \hat{\mathbf{s}}_R \cdot\left(\mathbf{a}_0^{(s)}+\frac{\alpha}{\pi}\mathbf{a}^{(s)} \right) \right],
    \label{total2}
\end{equation}

\noindent
where

\begin{equation}
a=\left[\text{Re}(\phi_n)+I_{0n}\right]a_0'+\text{Re}(\phi_n')a_V'+a_B'.
\label{a}
\end{equation}

\noindent
and

\begin{equation}
\mathbf{a}^{(s)}=\left[\text{Re}(\phi_n)+I_{0n}\right]\mathbf{a}_0^{(s)}+\text{Re}(\phi_n')\mathbf{a}_V^{(s)}+\mathbf{a}_B^{(s)}.
\label{as}
\end{equation}

\noindent
The vectors that constitute $\mathbf{a}^{(s)}$, namely $\mathbf{a}_0^{(s)}$, $\mathbf{a}_V^{(s)}$ and $\mathbf{a}_B^{(s)}$, arise from spin-dependent contributions to the decay amplitude. The terms $\text{Re}(\phi_n)+I_{0n}$ and $\text{Re}(\phi_n')$ are provided in Ref. \cite{2015}. ALthough $\phi_n$ and $I_{0n}$ encode separately infrared divergent terms, $\text{Re}(\phi_n)+I_{0n}$ is finite.

\noindent
On the other hand, the spin independent term $a$ in Eq. (\ref{a}) is written in terms of $a_0'$, which is defined in Eq. (\ref{a0p}) and

\begin{equation}
    a_V'=\frac{M_1^3}{8}\left[A_{1n}^{(V)} +A_{2n}^{(V)}Re\xi(q^2)+A_{3n}^{(V)}|\xi(q^2)|^2\right],
\end{equation}

\begin{equation}
    a_B'=\frac{M_1^3}{8}\left[A_{1n}^{(B)} +A_{2n}^{(B}Re\xi(q^2)+A_{3n}^{(B)}|\xi(q^2)|^2\right],
\end{equation}

\noindent
where $A_{jn}^{(V)}$ and $A_{jn}^{(B)}$, $j=1,2,3$, are defined in Ref. \cite{2015}.

\noindent
The triple integration over the real photon variables was performed numerically. However, the infrared divergence and the finite terms involved have been extracted analytically. 

\noindent
Although Eq. (\ref{total2}) is long in extension, is organized in a way that is easy to handle, thus, it can be used to evaluate the effects of RC on muon polarization, which will be described in the next section.

\section{The Muon Polarization Vector With RC}
\label{section:pola}

At this point, the results can be organized to construct the muon polarization vector with RC, hereafter denoted as $\mathbf{P}$

\begin{equation}
\mathbf{P}=\mathbf{P}_0+\mathbf{P}_{RC},
\label{polvec}
\end{equation}

\noindent
where $\mathbf{P}_0$ is the uncorrected muon polarization vector defined in Eq. (\ref{polarizacion0}) and $\mathbf{P}_{RC}$ contains both virtual and bremsstrahlung RC contributions to $\mathbf{P}_0$. It can be written as

\begin{equation}
\mathbf{P}_{RC}=\frac{\alpha}{\pi}\frac{\mathbf{a}^{(s)}-a\mathbf{P}_0}{a_0'+(\alpha / \pi)a},
\label{polrc}
\end{equation}

\noindent
where $\mathbf{a}^{(s)}$ and $a$ are defined in (\ref{as}) and (\ref{a}) respectively.

\noindent
The magnitude of the muon polarization vector is obtained in the usual way as


\begin{equation}
    P=|\mathbf{P}|=\sqrt{\left(P_{L0}+P_{LRC}\right)^2+\left(P_{T0}+P_{TRC}\right)^2+\left(P_{N0}+P_{NRC}\right)^2},
    \label{polmag}
\end{equation}

\noindent
where $P_{X0}$ and $P_{XRC}$, $X=L,T,N$, are the longitudinal, transverse, and normal components of $\mathbf{P}_0$ and $\mathbf{P}_{RC}$, respectively.

\noindent
With all the inputs, the numerical evaluation of the RC to the components of the muon polarization vector Eq. (\ref{polvec}), and its magnitude $P$, can be performed at any point in the allowed kinematical region. Samples of these numbers are shown in Tables \ref{tab:rc} and \ref{tab:tot} respectively.

\begin{table*}[ht]
\centering
\resizebox{15cm}{!} {
\begin{tabular}{c|rrrrrrrrrr}
\hline
 $E_2$\textbackslash$E$ &$0.1123$&$0.1258$&$0.1393$&$0.1528$&$0.1663$&$0.1797$&$0.1932$&$0.2067$&$0.2202$&$0.2337$\\ \hline
 
 (a)\\
        $0.2512$ & ~ & ~ & ~ & ~ & ~ & $0.0389$ & $0.0901$ & $0.1678$ & $0.2833$ & $0.4561$  \\ 
        
        $0.2395$ & ~ & $-0.0026$ & $-0.0130$ & $0.0098$ & $0.0525$ & $0.1110$ & $0.1846$ & $0.2738$ & $0.3760$ & $0.3778$ \\ 
        
        $0.2277$ & $-0.0408$ & $0.0705$ & $-0.0513$ & $-0.0079$ & $0.0513$ & $0.1228$ & $0.2045$ & $0.2936$ & $0.3733$ & $0.1508$  \\ 
        
        $0.2160$ & $-0.0246$ & $-0.0873$ & $-0.0722$ & $-0.0230$ & $0.0441$ & $0.1225$ & $0.2075$ & $0.2920$ & $0.3425$ & $-0.2805$  \\ 
        
        $0.2042$ & $0.2556$ & $-0.0312$ & $-0.0654$ & $-0.0270$ & $0.0404$ & $0.1208$ & $0.2052$ & $0.2807$ & $0.2907$ & $-1.0956$  \\ 
        
        $0.1924$ & ~ & $0.1668$ & $-0.0100$ & $-0.0100$ & $0.0464$ & $0.1228$ & $0.2018$ & $0.2616$ & $0.2079$ &   \\ 
        
        $0.1807$ & ~ & ~ & $0.1288$ & $0.0385$ & $0.0673$ & $0.1312$ & $0.1974$ & $0.2294$ & $0.0596$ &   \\ 
        
        $0.1689$ & ~ & ~ & ~ & $0.1079$ & $0.0962$ & $0.1400$ & $0.1828$ & $0.1629$ & $-0.2584$ &   \\ 
        
        $0.1572$ & ~ & ~ & ~ & ~ & $0.0292$ & $0.0967$ & $0.1116$ & $-0.0145$ & ~ &   \\ 
        
        $0.1454$ & ~ & ~ & ~ & ~ & ~ & $-0.3942$ & $-0.2548$ & $-0.5575$ & ~ &   \\
 
 (b)\\
     $0.2512$ & ~ & ~ & ~ & ~ & ~ & $-0.1057$ & $-0.1880$ & $-0.2382$ & $-0.2535$ & $-0.2236$  \\ 
        
        $0.2395$ & ~ & $-0.1286$ & $-0.3069$ & $-0.4215$ & $-0.4945$ & $-0.5318$ & $-0.5332$ & $-0.4927$ & $-0.3961$ & $-0.2623$ \\ 
        
        $0.2277$ & $-0.1748$ & $-0.3769$ & $-0.5073$ & $-0.5901$ & $-0.6305$ & $-0.6296$ & $-0.5842$ & $-0.4864$ & $-0.32494$ & $-0.2539$  \\ 
        
        $0.2160$ & $-0.1357$ & $-0.3235$ & $-0.4574$ & $-0.5380$ & $-0.5679$ & $-0.5478$ & $-0.4755$ & $-0.3456$ & $-0.1643$ & $-0.3698 $ \\ 
        
        $0.2042$ & $0.0744$ & $-0.0674$ & $-0.2194$ & $-0.3164$ & $-0.3535$ & $-0.3323$ & $-0.2534$ & $-0.1197$ & $0.0265$ & $-0.3290$  \\ 
        
        $0.1924$ & ~ & $0.3067$ & $0.1772$ & $0.0547$ & $-0.0078$ & $-0.0062$ & $0.0541$ & $0.1555$ & $0.1919$ &   \\ 
        
        $0.1807$ & ~ & ~ & $0.6944$ & $0.5633$ & $0.4572$ & $0.4121$ & $0.4200$ & $0.4387$ & $0.2547$ &   \\ 
        
        $0.1689$ & ~ & ~ & ~ & $1.1877$ & $1.0248$ & $0.8957$ &$ 0.8012 $& $0.6604$ & $0.0913$ &   \\ 
        
        $0.1572$ & ~ & ~ & ~ & ~ & $1.6462$ & $1.3749$ & $1.0971$ & $0.6691$  & ~ &   \\ 
        
        $0.1454$ & ~ & ~ & ~ & ~ & ~ & $1.6028$ & $0.9689$ & $0.1062$ & ~ &   \\
 
 (c)\\
     $0.2512$ & ~ & ~ & ~ & ~ & ~ & $-0.0061$ & $-0.0139$ & $-0.0285$ & $-0.0578$ & $-0.1669$ \\ 
        
        $0.2395$ & ~ & $-0.0003$ & $0.0071$ & $-0.0002$ & $-0.0152$ & $-0.0353$ & $-0.0613$ & $-0.0986$ & $-0.1672$ & $-0.4352$  \\ 
        
        $0.2277$ & $0.0401$ & $0.0667$ & $0.0433$ & $0.0105$ & $-0.0245$ & $-0.0616$ & $-0.1043$ & $-0.1623$ & $-0.2685$ & $-0.7195 $\\ 
        
        $0.2160$ & $0.1986$ & $0.1628$ & $0.0924$ & $0.0272$ & $-0.0312$ & $-0.0871$ & $-0.1476$ & $-0.2273$ & $-0.3744$ & $-1.0379$  \\ 
        
        $0.2042$ & $0.2475$ & $0.2752$ & $0.1520$ & $0.0476$ & $-0.0383$ & $-0.1153$ & $-0.1952$ & $-0.2986$ & $-0.4922$ & $-0.5859$ \\ 
        
        $0.1924$ & ~ & $0.3510$ & $0.2120$ & $0.0651$ & $-0.0519$ & $-0.1521$ & $-0.2527$ & $-0.3820$ & $-0.6289$ &   \\ 
        
        $0.1807$ & ~ & ~ & $0.2310$ & $0.0609$ & $-0.0857$& $-0.2083$ & $-0.3294$ & $-0.4857$ & $-0.7867$ &   \\ 
        
        $0.1689$ & ~ & ~ & ~ & $-0.0151$ & $-0.1722$ & $-0.3074$ & $-0.4424$ & $-0.6207$ & $-0.9121$ &   \\ 
        
        $0.1572$ & ~ & ~ & ~ & ~ & $-0.3687$ & $-0.4950$ & $-0.6177$ & $-0.7794$ & ~ &   \\ 
        
        $0.1454$ & ~ & ~ & ~ & ~ & ~ & $-0.6448$ & $-0.7731$ & $-0.5957$ & ~ &   \\
\hline  
\end{tabular}
}
\caption{RC to the components of the muon polarization $\mathbf{P}_0$. Eq. (\ref{polrc}), in the TBR of process $K_{\mu 3}^0$. The entries correspond to (a)$P_{LRC}\times 10^{2}$, 
 (b)$P_{TRC}\times 10^{4}$, and (c)$P_{NRC}$. The energies $E$ and $E_2$ are given in GeV. For definiteness, $\text{Re}\xi(0)=-0.151$ and $\text{Im}\xi(0)=-0.007$ are used. \label{tab:rc}}
\end{table*}

\begin{table*}[ht]
\centering
\resizebox{15cm}{!} {
\begin{tabular}{c|rrrrrrrrrr}
\hline
$E_2$\textbackslash$E$ &$0.1123$&$0.1258$&$0.1393$&$0.1528$&$0.1663$&$0.1797$&$0.1932$&$0.2067$&$0.2202$&$0.2337$\\ \hline

        $0.2512$ & ~ & ~ & ~ & ~ & ~ & $1.0004$ & $1.0009$ & $1.0017$ & $1.0029$ & $1.0048$  \\ 
        
        $0.2395$ & ~ & $1.0000$ & $0.9999$ & $1.0001$ & $1.0006$ & $1.0012$ & $1.0019$ & $1.0029$ & $1.0041$ & $1.0054$ \\ 
        
        $0.2277$ & $0.9994$ & $0.9990$& $0.9993$ & $0.9999$ & $1.0006$ & $1.0014$ & $1.0023$ & $1.0033$ & $1.0045$ & $1.0060$ \\ 
        
        $0.2160$ & $0.9980$ & $0.9982$ & $0.9988$ & $0.9997$ & $1.0005$ & $1.0015$ & $1.0025$ & $1.0036$ & $1.0049$ & $1.0083$  \\ 
        
        $0.2042$ & $0.9964$ & $0.9973$ & $0.9984$ & $0.9995$ & $1.0006$ & $1.0017$ & $1.0028$ & $1.0040$ & $1.0053$ & $1.0122$ \\ 
        
        $0.1924$ & ~ & $0.9962$ & $0.9980$ & $0.9994$ & $1.0007$ & $1.0019$ & $1.0032$ & $1.0045$ & $1.0061$ &   \\ 
        
        $0.1807$ & ~ & ~ & $0.9975$ & $0.9995$ & $1.0011$ & $1.0025$ & $1.0038$ & $1.0052$ & $1.0072$ &   \\ 
        
        $0.1689$ & ~ & ~ & ~ &$ 1.0000$ & $1.0019$ & $1.0034$ & $1.0047$ & $1.0062$ & $1.0089$ &   \\ 
        
        $0.1572$ & ~ & ~ & ~ & ~ & $1.0035$ & $1.0050$ & $1.0062$ & $1.0077$ & ~ &   \\ 
        
        $0.1454$ & ~ & ~ & ~ & ~ & ~ & $1.0075$ & $1.0081$ & $1.0081$ & ~ &   \\
\hline      
\end{tabular}
}
\caption{The magnitude of the muon polarization vector $P$, Eq. (\ref{polmag}), in the TBR of process $K_{\mu 3}^0$. The energies $E$ and $E_2$ are given in GeV. For definiteness, $\text{Re}\xi(0)=-0.151$ and $\text{Im}\xi(0)=-0.007$ are used.\label{tab:tot} }
\end{table*}

\noindent
A final step of integration over $E$ and $E_2$ can be performed on Eq. (\ref{polvec}) to obtain the totally integrated components of the muon polarization $P_X^{tot}$. With all the necessary inputs, the numerical values are found to be

\begin{subequations}
    \begin{align}
        P_L^{tot}=0.7101-0.0038,\\
        \intertext{}
        P_T^{tot}=-0.0015+0.0020\times 10^{-3},\\
        \intertext{and}
        P_N^{tot}=-0.5391-0.0017,
    \end{align}
   \label{pfully} 
\end{subequations}

\noindent
where the first term in each of the equations above is the uncorrected value and the second term is the RC from TBR. In particular, the $P_T^{tot}$ value is of the order of magnitude of the measured value suggested by the literature.

\section{Conclusions and Remarks}
\label{section:con}

In this paper we have obtained the radiative corrections to the Dalitz plot of $K_{\mu 3}^0$ decays to order $(\alpha / \pi )(q/M_1)$, given by Eq. (\ref{total}), where $q$ is the momentum transfer and $M_1$ denotes the mass of kaon. This result can be written in the compact form as shown in Eq. (\ref{total2}). This expression comprises contributions of both virtual and bremsstrahlung RC restricted to the three-body part of the allowed kinematical region. Despite its length, Eq. (\ref{total2}) it is simple and organized. It also has other properties, it contains all the terms of order $(\alpha / \pi )(q/M_1)$, does not have an infrared divergence or ultraviolet cutoff and it is not compromised by any model dependence of radiative corrections.

The advantage of Eq. (\ref{total2}) lies in its simplicity for evaluating several physical observables such as the muon polarization vector.

From $d\Gamma(K_{\mu 3}^0)$, Eq. (\ref{total2}), the muon polarization vector $\mathbf{P}$ in the variables $E$ and $E_2$ with RC can be obtained. It is given by Eq. (\ref{polvec}).

As a first result, in Table \ref{tab:0}, we present the components of the uncorrected muon polarization vector $\mathbf{P}_0$. Relative to the longitudinal and normal directions, the values of the transverse component, namely $P_{T0}$, are two orders of magnitude below.

The values of RC to the components of the uncorrected muon polarization $\mathbf{P}_0$ corresponding to the three body region, are listed in Table \ref{tab:rc}, namely $P_L$, $P_T$ and $P_N$. Relative to the uncorrected values, at the longitudinal component $P_{LRC}$ the lowest percentage are approximately $0.002\%$ for the muon's energies closest to its minimum, and the highest percentage reaches approximately $1.41\%$ for energies near the maximum energy of the muon. A similar behavior is observed in the normal component $P_{NRC}$, with $0.001\%$ as the minimum and $1.14\%$ as the higher percentage reached at the same regions of the DP. For the transverse component, the values of the RC, namely $P_{TRC}$ are approximately $0.027\%$ at the central region of the DP, whereas the maximum percentage reached is approximately $6\%$ for pion's energies at its minimum and maximum in the same region. These conclusions are reached for the values $\text{Re}\xi(0)=-0.151$ and $\text{Im}\xi(0)=-0.007$ which are closest to physics. Other values might take into different scenarios.

Additionally, the magnitude of the polarization vector $P$, Eq (\ref{polmag}), evaluated at several points of the DP is presented in Table \ref{tab:tot}. The highest correction reached on $P$ is approximately $1\%$ at the lower-right corner of the DP. Although the maximum percentages reached at the transverse component are relatively higher compared with the other two components, the magnitude of the polarization vector does not seem to be affected because the values for $P_{TRC}$ are very small, approximately two orders of magnitude below relative to the other two components.

Finally, the fully integrated components of the polarization in the TBR Eq. (\ref{pfully}) get corrections of $0.5\%$ and $0.3\%$ for the longitudinal and normal components and $1.4\%$ for the transverse component.

To complete the analysis, an extra effort will be made to treat the so-called four body region of the DP.

\acknowledgments

The authors are grateful to Consejo Nacional de Ciencia y Tecnología (México) for partial support.


\bibliography{biblio}


\end{document}